\documentclass[pre,aps,superscriptaddress,showpacs]{revtex4-1}
\usepackage{natbib}
\usepackage{amsmath}
\usepackage{epsfig}

\begin{document}

\title{A time-work uncertainty relation in quantum systems}
\author{Fei Liu}
\email[Email address: ]{feiliu@buaa.edu.cn}
\affiliation{School of Physics and Nuclear Energy Engineering, Beihang University, Beijing 100191, China}
\date{\today}

\begin{abstract}
{In quantum systems, a plausible definition of work is based on two energy measurement scheme. Considering that energy change of quantum system obeys a time-energy uncertainty relation, it shall be interesting to see whether such type of work as well obeys an analogous uncertainty relation. In this note I argue that this relation indeed exists for closed quantum systems and open quantum systems, which the latter are assumed to be weakly coupled with their environments. }

\end{abstract}
\maketitle

In the two past decades, there has been growing interest in stochastic work of nonequilibrium quantum processes~\cite{Kurchan2000,Tasaki2000,Allahverdyan2005,Talkner2007,Esposito2009,Campisi2011}. In order to formulate a quantum version of the celebrated Jarzynski equality~\cite{Jarzynski1997}, in a closed quantum system, a two-energy measurements (TEM) scheme was proposed by Kurchan~\cite{Kurchan2000} to define the work. Assume that the Hamiltonian of a closed quantum system is $H(t)$ and $\Delta t$ is the duration of the nonequilibrium quantum process, that is, $\Delta t=t_f-t_0$, where $t_0$ and $t_f$ are the initial and end times, respectively. Let $E_0$ be the initial energy of the quantum system, i.e., one of eigenvalues of $H(t_0)$ at the initial time. $E$ is the measured energy of the system with $H(t_f)$ at the end time. Note that this $E$ is implicitly influenced by $E_0$, since the system starts from the eigenvector of $H(t_0)$ with the eigenvalue $E_0$. Then the TEM defines the work of this process to be
\begin{eqnarray}
\label{workenergy}
W= E-E_0.
\end{eqnarray}
This work is stochastic due to the randomness of $E$. The latter can be described by a distribution $p_{E_0}(E)$ that is the probability of finding the system in the eigenvector of $H(t_f)$ with the eigenvalue $E$. On the other hand, the quantum nature of this process also implies a time-energy uncertainty relation~\cite{LandauQuantum,Busch2007}
\begin{eqnarray}
\label{TEUR}
\sigma_{E|E_0} \Delta t \ge \hbar,
\end{eqnarray}
where the variance of energy is
\begin{eqnarray}
\sigma^2_{E|E_0}=\int p_{E_0}(E) \left(E-\overline{E|E_0}\right)^2 dE,
\end{eqnarray}
and $\overline{E|E_0}$ is the mean of measured energy, that is,
\begin{eqnarray}
\overline{E|E_0}=\int p_{E_0}(E) E dE.
\end{eqnarray}
According to Eq.~(\ref{workenergy}), if the distribution of the work is $\rho_{E_0}(W)$, one must have $\rho_{E_0}(W)dW=p(E|E_0)dE$. Therefore, the mean and variance of the work are
\begin{eqnarray}
\overline{W|E_0}=\int \rho_{E_0}(W) W dW=\int p_{E_0}(E) (E-E_0) dE=\overline {E|E_0} -E_0,
\end{eqnarray}
and
\begin{eqnarray}
\sigma^2_{W|E_0}=\int \rho_{E_0}(W) \left(W-\overline{W|E_0}\right)^2 dW=\sigma^2_{E|E_0},
\end{eqnarray}
respectively. Based on the time-energy uncertainty relation~(\ref{TEUR}), I then find a time-work uncertainty relation under a fixed initial energy $E_0$:
\begin{eqnarray}
\label{TWURE0}
\sigma_{W|E_0}\Delta t\ge \hbar.
\end{eqnarray}
In general, the initial energy $E_0$ is also a random number, that is, it may has a distribution, e.g., the conventional canonical distribution~\cite{Campisi2011}. Hence, I need to extend the above relation into the mixed initial state. This is not difficult if the reader recalls the law of total variance,
\begin{eqnarray}
\sigma^2_{W}&=&\langle \sigma^2_{W|E_0} \rangle_{E_0}+ \sigma^2_{ \overline{W|E_0}},
\end{eqnarray}
where these two terms are the mean of $\sigma^2_{W|E_0}$ and the variance of $\overline{W|E_0}$ respect to the randomness of $E_0$, respectively. Since the latter term is always positive, I immediately get that
\begin{eqnarray}
\label{TWUR}
\sigma_{W} \Delta t \ge \hbar.
\end{eqnarray}

Because closed quantum systems are not common in reality, it shall be interesting to investigate the uncertainty relation~(\ref{TWUR}) in the open quantum regime~\cite{Breuer2002}. Here I restrict in the case that the system is weakly coupled with an environment and external driving directly acts on the system. Following the idea of Talkner et al.~\cite{Talkner2009}, I first regard the system and its environment as a composite system. The total Hamiltonian of the composite system is
\begin{eqnarray}\label{totalHamiltonian}
H(t)=H_s(t)+H_e+H_i,
\end{eqnarray}
where these three terms are the Hamiltonian of the system, the environment, and their interaction, respectively.  To define work, the TEM scheme is simultaneously conducted on the system and environment at the beginning and the end of the nonequilibrium quantum process, respectively. The measured energy change for the system is referred to as the internal energy change of system, while the energy change for the environment is referred to as the heat released from the system. Then, the work done on the open system is defined as the sum of the internal energy change and the heat. Nevertheless, under the assumption of the weak interaction between these two systems, this work definition of the open quantum system is nothing but the work defined on the closed composite system~(\ref{totalHamiltonian}). Of course, the latter obeys Eq.~(\ref{TWUR}). Therefore, I conclude that the time-work uncertainty relation is still valid even in these quantum open systems.


I close this note by presenting several comments on Eq.~(\ref{TWUR}). First, there exist a variety of time-energy uncertainty relations. Validity and meaning of them are often very controversial in the literature~\cite{Busch2007}. Nevertheless, Eq.~(\ref{TEUR}) is one that is widely accepted. Second, the work definition using the TEM scheme is also controversial by the fact that the scheme may destroy the initial quantum-coherent superposition of quantum system~\cite{Allahverdyan2005}. Therefore, it would be inevitable that the time-work uncertainty relation~(\ref{TWUR}) has to be faced the same criticism. Finally, it shall be important to verify this relation by using concrete quantum models, e.g., rapidly expanding quantum piston model by Quan and Jarzynski~\cite{Quan2012} and weakly driven two-level dissipative quantum system~\cite{Liu2014}.\\
\\

{\noindent \it Acknowledgment.} I thank Professor Haitao Quan for his useful remarks on the note. The work was supported by the National Science Foundations of China under Grant No. 11174025 and No. 11575016.

\bibliography{RFsubmission20160729}.
\end{document}